\documentclass{article}
\pdfoutput=1
\usepackage{graphics}
\usepackage[letterpaper,portrait]{geometry}
\usepackage{amssymb}
\usepackage{chapterbib}
\usepackage[super,sort&compress]{natbib}
\title {High-resolution spectroscopy of two-dimensional electron systems}
\author {O. E. Dial\footnote{Massachusetts Institute of Technology, Cambridge, MA 02139, USA}, R. C. Ashoori\footnotemark[1], L. N. Pfeiffer\footnote{Bell Laboratories, Alcatel-Lucent, Murray Hill, New Jersey 07974, USA}, K. W. West\footnotemark[2]}
\begin{document}
\begin{cbunit}
\maketitle
\begin{abstract}
Spectroscopic methods involving the sudden injection or ejection of
electrons in materials are a powerful probe of electronic structure
and interactions\cite{Giaever60}. These techniques, such as
photoemission and tunneling, yield measurements of the ``single
particle'' density of states (SPDOS) spectrum of a
system\cite{Bardeen61}.  The SPDOS is proportional to the probability
of successfully injecting or ejecting an electron in these
experiments.  It is equal to the number of electronic states in the
system able to accept an injected electron as a function of its energy
and is among the most fundamental and directly calculable quantities
in theories of highly interacting \mbox{systems\cite{Hedin71}}.
However, the two-dimensional electron system (2DES), host to
remarkable correlated electron states such as the fractional quantum
Hall effect\cite{Stormer99}, has proven difficult to probe
spectroscopically.  Here we present an improved version of time domain
capacitance spectroscopy \mbox{(TDCS)\cite{Chan97}} that now allows
us to measure the SPDOS of a 2DES with unprecedented fidelity and
resolution.  Using TDCS, we perform measurements of a cold 2DES,
providing the first direct measurements of the single-particle
exchange-enhanced spin gap\cite{Ando74} and single particle
lifetimes\cite{Chaplik71a} in the quantum Hall system, as well as the
first observations of exchange splitting of Landau levels not at the
Fermi surface.  The measurements reveal the difficult to reach and
beautiful structure present in this highly correlated system far from
the Fermi surface.
\end{abstract}
Spectroscopy of the 2DES has been performed optically and using
electron tunneling.  Traditional optical methods probe long-wavelength
collective excitations rather than the SPDOS\cite{Pinczuk88}.
Attempts to measure the SPDOS optically have relied on the
introduction of local impurities, altering the measured
spectrum\cite{Kukushkin96}.  Approaches based on tunneling
spectroscopy have been hampered by several
difficulties\cite{Mendez86,Eisenstein91,Main00}.  Typically, the
tunnel current enters the 2DES perpendicularly to the plane through a
tunnel barrier.  In steady-state measurements, maintenance of this
current requires it to have a path to exit the 2DES.  The current
can either flow in the plane of the 2DES to a distant contact, or
it can tunnel out perpendicularly from the other side of the plane.
However, if the in-plane conductivity of the device is small (as
frequently occurs in quantum Hall systems), the current cannot flow
out through the plane of the 2DES.  If the current tunnels out
perpendicularly through a second barrier, the two barriers form a
resistive divider and variations in the tunnel current alter the 2D
density and tunnel voltage.  Finally, ohmic heating limits the useful
range of both these techniques to small tunnel voltages.

TDCS eliminates all of the major problems associated with 2D tunneling
spectroscopy by abandoning steady-state measurement.  In TDCS, we
replace the ohmic contact to the 2DES with a distant, electrically
isolated electrode parallel to the plane of the 2DES.  We capacitively
apply the tunnel voltage using the isolated electrode and then
capacitively detect the tunneled electrons rather than removing them
through a second contact or barrier.  Because the electron does not
need to move in the plane of the 2DES to be detected, we can measure
the SPDOS even when the 2DES is insulating or empty.  Although this
capacitive coupling requires us to use a pulsed AC measurement to
extract I-V characteristics, it also allows us to use very short duty
cycles (typically 0.01\%) to minimize heating.

In our measurements, the 2DES is sandwiched between two electrodes of
a ``tunnel capacitor'' created from a GaAs heterostructure (Figure
\ref{fig:device}a,b). The tunneling electrode, used as a source for
injected electrons, is separated from the 2DES by a thin
$\mathrm{Al_{.33}Ga_{.67}As}$ tunneling barrier.  The isolated
electrode, used to apply voltages and detect charges, is separated by
a thicker, insulating barrier.  To minimize scattering, there is no
intentional doping in either barrier.  The 2D electron density is
controlled from 0 up to $4\times{}10^{11}/\mathrm{cm}^{2}$ by varying
the DC voltage applied across the electrodes.

To perform our measurement, we apply a sudden voltage step
$\mathrm{V}_\mathrm{P}$ ($\lesssim$1 ns rise time), disequilibrating the
tunneling electrode from the 2DES.  In the absence of bridge
compensation, charging the geometric capacitance ($\sim{}10$ pF)
between the electrodes would generate a large charge step at the
detector.  We remove this step by applying a pulse of opposite
polarity to a standard capacitor $\mathrm{C_{std}}$ connected to the isolated
electrode (Figure \ref{fig:device}c). As electrons tunnel across the
barrier, we measure a long slow rise in the total charge on the
isolated electrode due to their image charge.  Because electrons
tunnel only a short distance into the tunnel capacitor to reach the
2DES, this charging signal is much smaller ($<$~33\%) than the
uncompensated initial charge step.  The charging signal is measured
using a cryogenic amplifier and massive (250,000x) repetitive signal
averaging.  We fit the derivative of this signal as a function of time
to determine the tunneling current (Figure \ref{fig:device}d). In the
first instant after the step, the voltage V between the 2DES and the
tunneling electrode is precisely $\mathrm{V_P}$ multiplied by an
easily calibrated geometric lever-arm. This, together with the
measured tunneling current, gives us a single point on an I-V curve;
varying the pulse amplitude traces out the entire curve (Figure
\ref{fig:device}e).

With continuous application of the pulsed voltage, the 2DES and
electrode would re-equilibrate in a few milliseconds.  However, we use
only the first $\sim$100 ns of the step to determine the current,
before the voltage across the tunnel barrier has relaxed
significantly.  This short time interval means that little charge is
transferred, so electrons are tunneling into and out of cold
electronic systems.  After this brief measurement window has elapsed,
the voltage step is removed in preparation for the next pulse, and the
system is allowed to re-equilibrate.

Improving on our prior work\cite{Chan97}, with precise control over
the shape of the applied pulses, we are now able to reduce the
systematic and random errors in this I-V curve enough to numerically
differentiate it, giving $\mathrm{dI}/\mathrm{dV}$, a quantity
proportional to the SPDOS multiplied by a slowly varying tunneling
matrix element (see supplement).  Although the energy
resolution is fundamentally limited only by temperature (see the inset
of Figure \ref{fig:lifetime}a), the colorscale plots are taken with
lower resolution to reduce acquisition time.

Figure \ref{fig:dIdV} shows typical plots of the SPDOS as a function
of energy and Landau level filling fraction $\nu$.  We extract the
value of the geometric lever-arm by measuring the known cyclotron
splitting of the empty well.  We estimate the maximum error that
accrues in the energy calibration (the error in the $y$-axis) as we
move away from zero density and energy to be about 1\%, predominately
due to variation of the lever-arm from polarization of the 2DES and
depletion of the electrodes.

From left to right within each plot, the quantum well starts fully
depleted, with Landau levels visible above the Fermi energy ($E>0$).
As the well begins to fill, a magnetic field induced Coulomb gap
\cite{Chan97,Aleiner95b,Ashoori90,Eisenstein92}, appearing as a dark
band across E=0 in each plot, opens at the Fermi surface.
Each Landau level is pinned by its high density of states as it
crosses the Fermi energy.  This results in a stair-step pattern of
successive flat pinned regions followed by downward steps
corresponding to chemical potential jumps as the next Landau level is
pulled down to the Fermi energy\cite{Popov06}.

Exchange-enhanced spin-splitting is visible as splitting of the Landau
levels as they cross the Fermi surface.  At even-integer filling,
there is no net spin polarization, causing the exchange gap to vanish.
As the density is raised, one spin state of the orbital Landau level
begins to fill, and the attractive exchange interaction creates a gap
to the other spin state of the orbital level that grows until it
saturates with one completely filled spin state at odd-integer $\nu$.
The gap shrinks again as the upper spin state is filled and pulled
down in energy to meet the lower spin state.  This gives rise to the
``oscillatory g-factor'' in quantum Hall systems\cite{Ando74} and, 
together with the pinning of the spin state being filled to the
Fermi energy, creates a distinctive pattern as each Landau
level crosses the Fermi surface.  

We measure the spin-splittings, $E_j$, at odd-integer filling by
fitting the peaks in constant $\nu$ line-cuts of the
spectra to Lorentzians and develop an empirical formula for
describing the splittings.  The results are shown in Figure
\ref{fig:exchange}a.  Varying density at fixed values of the magnetic
field, we find that the spin gaps diminish with a $1/\sqrt{\nu}$
dependence. To compare the B-field dependence of splittings at
different $\nu$, Figure \ref{fig:exchange}c shows the
splittings multiplied by $\sqrt{\nu}$ as a function of B-field. The
data cleanly collapse onto a single line, revealing a B-linear
dependence. The splittings are well
characterized by $E_j{}={}1.2{}\pm{}0.03\ {\mathrm{B}}/{\sqrt{\nu}}
~{\mathrm{meV}}/{\mathrm{T}}$ (Figure \ref{fig:exchange}b), where we
have forced the fit through the origin.

No previous experiment has directly measured the single-particle
exchange gap that arises in the sudden introduction of a single
electron. However, an exchange enhanced gap is also present in the
``thermodynamic'' density of states often measured using capacitance
measurements. These measurements monitor the change in the 2DES
chemical potential long after adding electrons to the system, allowing
time for correlations to form\cite{Dolgopolov97,Ashoori91}.  

Nonetheless, a B-linear dependence has also been seen both in
capacitance measurements of the gap in the thermodynamic density of
states\cite{Dolgopolov97} and in thermally activated measurements of
the transport gap\cite{Usher90}. However, the measured values of $E_j$
are roughly a factor of 4 or 2 smaller than our measurements,
respectively.  In contrast with our results, thermal activation
experiments find no $\nu$ dependence at fixed field\cite{Schmeller95,
Usher90}.

In the absence of Landau level mixing, the exchange-enhanced spin-gap
at the Fermi energy, $E_j$, is calculated\cite{MacDonald86,Aleiner95a}
to grow as $\sqrt{B}$ at fixed $\nu$ and to be largest at $\nu=1$
compared with other odd-integer fillings. However, we expect that
mixing is important; for $B$~$<$~6.5~T, the exchange energy
scale $e^{2}/\epsilon l_b$ is greater than the cyclotron splitting
$\hbar\omega_c$, and level repulsion from nearby Landau levels should
limit the exchange splitting to $\sim\hbar\omega_c$.  This B-linear
dependence has been found in structure-specific
calculations\cite{Smith92}.  More recent theory by Iordanski,
providing results only at $\nu=1$, predicts a B-linear splitting
roughly 30\% smaller than our value\cite{Iordanski02}.  No theory
predicts the observed $B/\sqrt{\nu}$ dependence.  

We also observe for the first time the indirect exchange splitting of
Landau levels above and below the Fermi surface by spin-polarized
Landau levels at the Fermi surface (yellow ovals in Figure
\ref{fig:dIdV}).  Calculations of the exchange matrix element
\cite{MacDonald86} between different orbital Landau levels indicate
that because of smaller wavefunction overlap, at any given density
this indirect splitting should be smaller than the exchange-enhanced
spin gap at the Fermi energy (neglecting Landau level mixing).  While
we find that this is sometimes correct, the most deeply buried orbital
Landau level $N=0$ is consistently split more than any other level in
the spectrum.  This level is unique in that there is no deeper Landau
level available for level mixing, so there is no level repulsion to
limit the exchange spitting on the lower energy side.  We suggest that
level mixing generally reduces these indirect exchange splittings just
as it reduces the splitting at the Fermi surface, but in the case of
$N=0$ the absence of lower energy Landau levels allows the splitting
to grow larger.

The most remarkable aspect of this observation is that the split
Landau levels are at energies several hundred times the thermal
energy, yet their splitting reflects the relatively delicate exchange
splitting at the Fermi surface.  This demonstrates that the high
energy features in the spectra depend on properties of the many-body
system only present at low temperatures. The high energy
spectra constitute a new and previously unexplored probe of the highly
correlated many-body ground state.

The lineshapes of Landau levels away from the Fermi energy provide a
unique window into the effects of electron-electron interactions.  We
find that the lineshape at fields of 1 Tesla or less fits well to
Lorentzians (inset, Figure \ref{fig:lifetime}a).  With a nearly empty
quantum well, sharp Landau levels are observed at any energy
(blue box in Figure \ref{fig:lifetime}a).  As the 2DES is filled, the
levels far from the Fermi surface suddenly broaden and disappear.  In
an occupied 2DES, the injected electron can scatter off electrons
already present in the Fermi sea.  The phase space available for this
scattering grows rapidly as the injection energy moves away from the
Fermi energy.  At large energies, the short lifetime due to
electron-electron scattering becomes the dominant source of
broadening.  By comparison, at low energies we find maxima in the peak
width at even-integer $\nu$ and minima at odd-integer $\nu$.  Here,
disorder is the chief source of broadening.  As observed, this
disorder induced broadening should display maxima at even-integer
$\nu$, where the density of states of the 2DES at the Fermi level is
small and the 2DES screens poorly.  It shows minima elsewhere, where the
density of states is high and the 2DES screens disorder well.

In the case of a disorder-free 2DES, the electron lifetime $\tau_{ee}$
has been calculated at zero field by Chaplik (see
supplement)\cite{Chaplik71a}.  At small magnetic fields such that the
cyclotron energy is small compared to the Coulomb interaction energy
we expect this result to hold.  Fitting the widths of the Landau
levels to Lorentzians gives us an irregularly spaced mesh of
broadenings as a function of energy and density.  In order to simplify
comparisons, we fit a smooth function through the observed width of
the Landau levels at each density, and plot contours of constant
broadening (Figure \ref{fig:lifetime}b).  At high energies and away
from even-integer $\nu$, we see good agreement with Chaplik's result
over a broad range of densities with no adjustable parameters; this is
the region where disorder broadening is small and we expect lifetime
broadening to dominate.  Prior measurements of momentum and phase
scattering times have found quantitative agreement with Chaplik's
result using an adjustable scaling parameter, but these data represent
the first direct measurement of the single particle
lifetime\cite{Murphy95,Yacoby91}.

\bibliographystyle{naturemag} \bibliography{article}
\vspace{.2in}

\hspace{0in}\textbf{Acknowledgements}\ \ This work was supported by Office of Naval
Research and by the National Science Foundation funded through the NSEC Program
and the MRSEC Program.

\clearpage
\begin{figure}
\center\resizebox{6in}{!}{\includegraphics{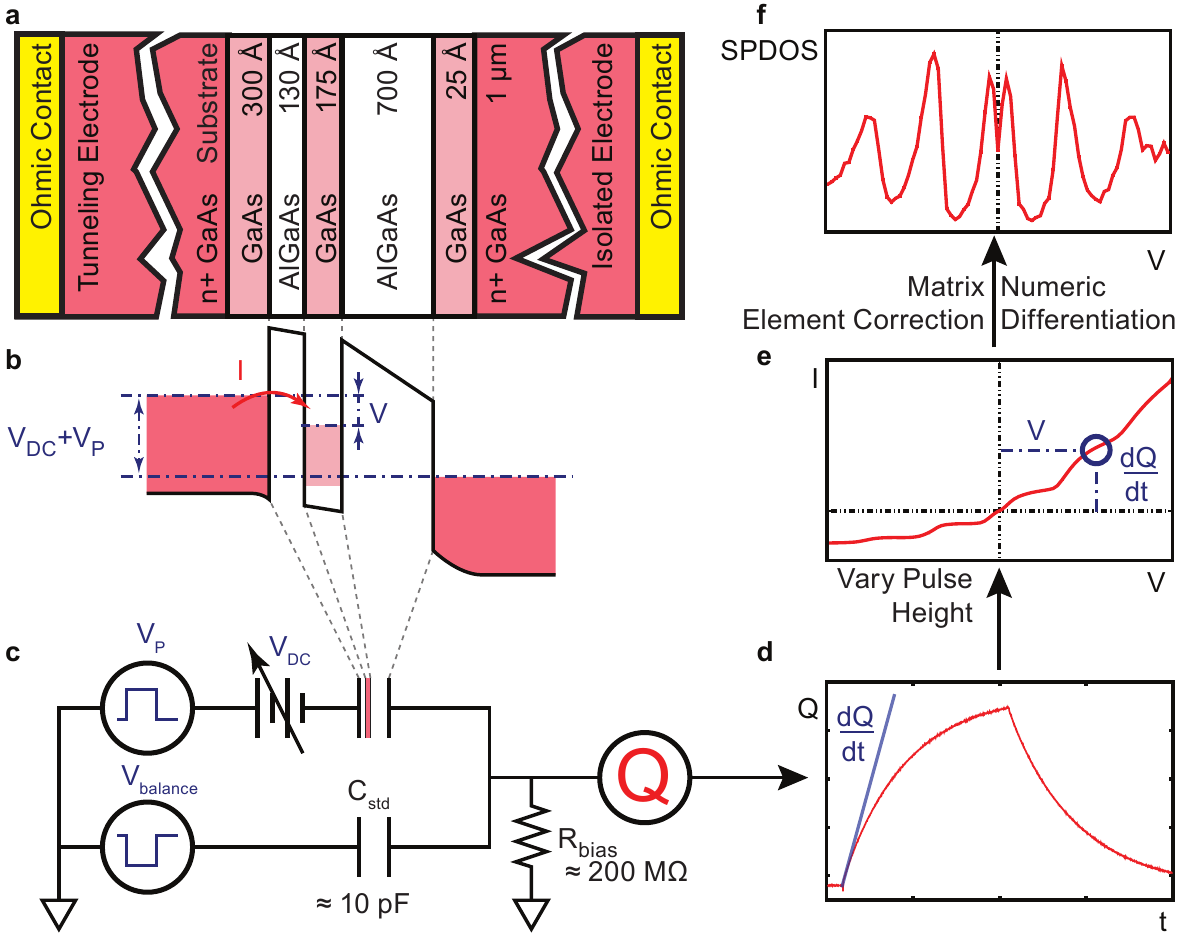}}
\caption{{\bf Time domain capacitance spectroscopy.} {\bf a}, GaAs
heterostructure used to construct the tunnel capacitor.  The only
ohmic contacts are to the top and bottom 3D doped regions, used
as electrodes.  {\bf b}, Band
structure of the tunnel capacitor in the instant after a pulse is
applied across it, showing the well defined voltage between the
tunneling electrode and 2DES, and the tunnel current.  {\bf c},
Simplified schematic of circuitry used to perform the measurement,
including a standard capacitor $\mathrm{C_{std}}$ used to remove the large
charge step from the background geometric capacitance of the
structure.  {\bf d}, Measured charge traces, taken with a long pulse
for illustrative purposes, showing the charge response averaged over
250,000 repetitions.  The slope of this curve immediately after the
pulse is applied gives the tunnel current.  {\bf e}, I-V characteristic
for tunneling into the 2DES constructed by combining data from $\sim$
200 different applied pulse heights.  {\bf f}, Single particle density
of states spectrum attained by multiplying {\bf e} by a smooth matrix
element correction and differentiating.}
\label{fig:device}
\end{figure}

\begin{figure}
\center\resizebox{6in}{!}{\includegraphics{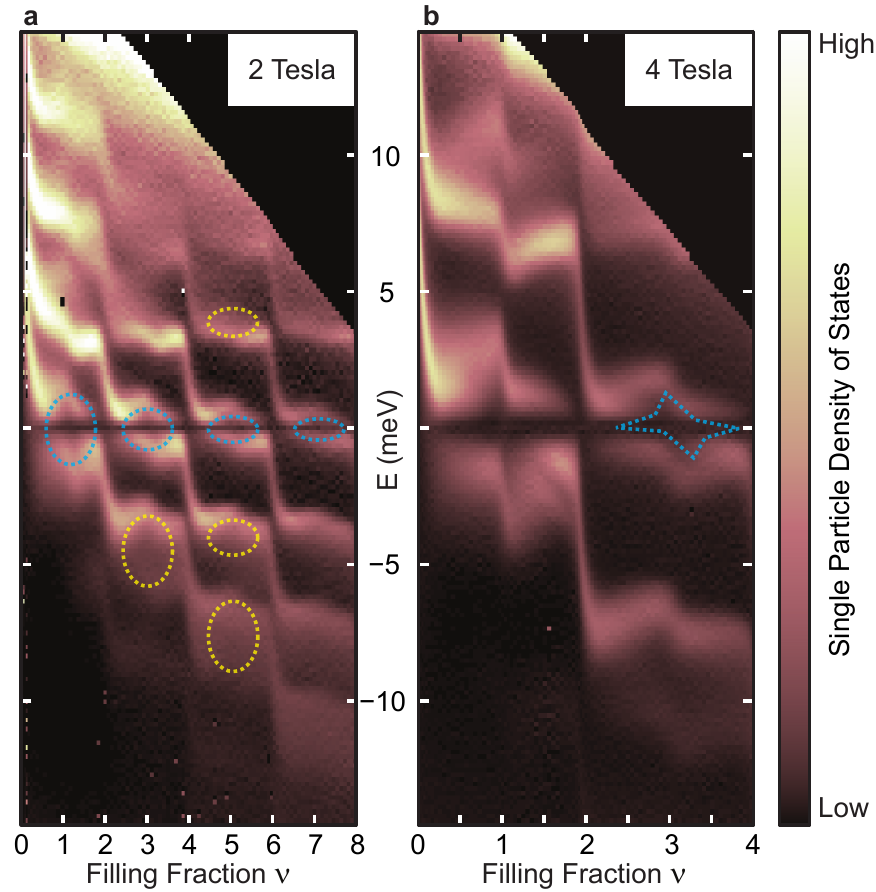}}
\caption{{\bf Single particle density of states spectra.} The spectra
are taken at fixed magnetic fields of 2 and 4 Tesla, with an energy resolution of 170 $\mu$eV and taken at
approximately 100 mK. Each bright peak corresponds to a Landau level.
The vertical ($y$) axis is the tunneling energy measured with respect
to the Fermi energy.  Energies greater than zero correspond to
injecting electrons into empty states in the quantum well, while
energies less than zero correspond to ejecting electrons from filled
states.  A smooth approximate correction for variation of the
tunneling matrix element has been applied (see supplement), resulting
in a dark band in the top-right corner of each plot.  The horizontal
($x$) axis is filling fraction $\nu$ (proportional to electron density
at fixed magnetic field).  The applied DC gate voltage is transformed
to $\nu$ through AC capacitance measurements\cite{Ashoori91}.  Blue
ellipses in the 2T plot mark exchange enhanced spin splittings, while
yellow ellipses mark some of the indirect exchange splittings.  The
blue dashed shape in the 4T plot shows the expected behavior of the
Landau levels as they cross the Fermi energy.}
\label{fig:dIdV}
\end{figure}

\begin{figure}
\center\resizebox{6in}{!}{\includegraphics{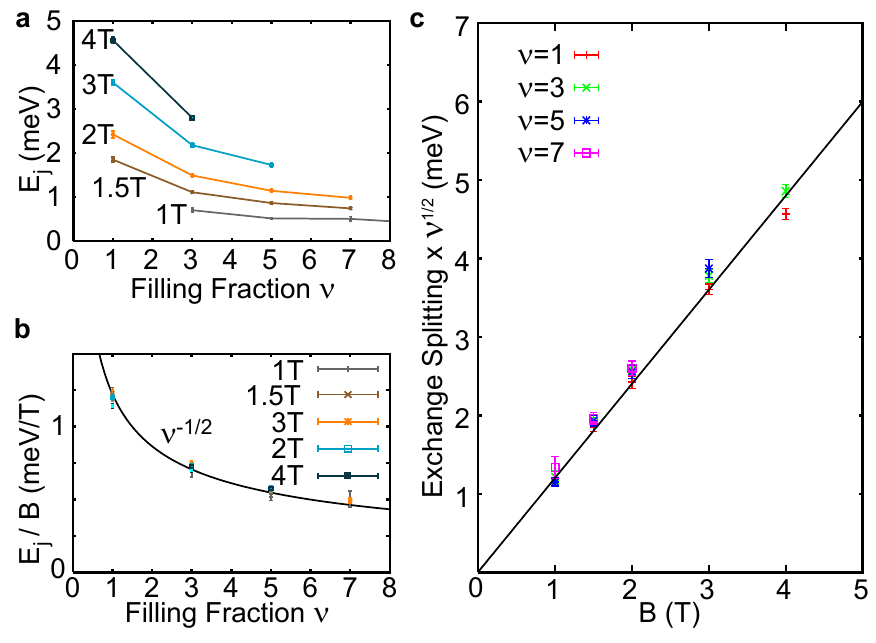}}
\caption{{\bf Exchange-enhanced spin splitting at the Fermi energy.}
{\bf a}, the unscaled exchange-enhanced spin gap as a function of odd-integer $\nu$ and field.
{\bf b}, $\nu^{-1/2}$ dependence (solid curve) shown by scaling the
splitting as a function of $\nu$ by 1/B.
{\bf c}, $E_j$ scaled by $\sqrt{\nu}$, showing the proportionality 
of $E_j$ to $\mathrm{B}/\sqrt{\nu}$.  The Landau levels are fit by a
superposition of Lorentzians to extract the splittings.  Error bars
show the standard deviation of the exchange splitting, as determined
by the variation of the fit residuals. }
\label{fig:exchange}
\end{figure}

\begin{figure}
\center\resizebox{6in}{!}{\includegraphics{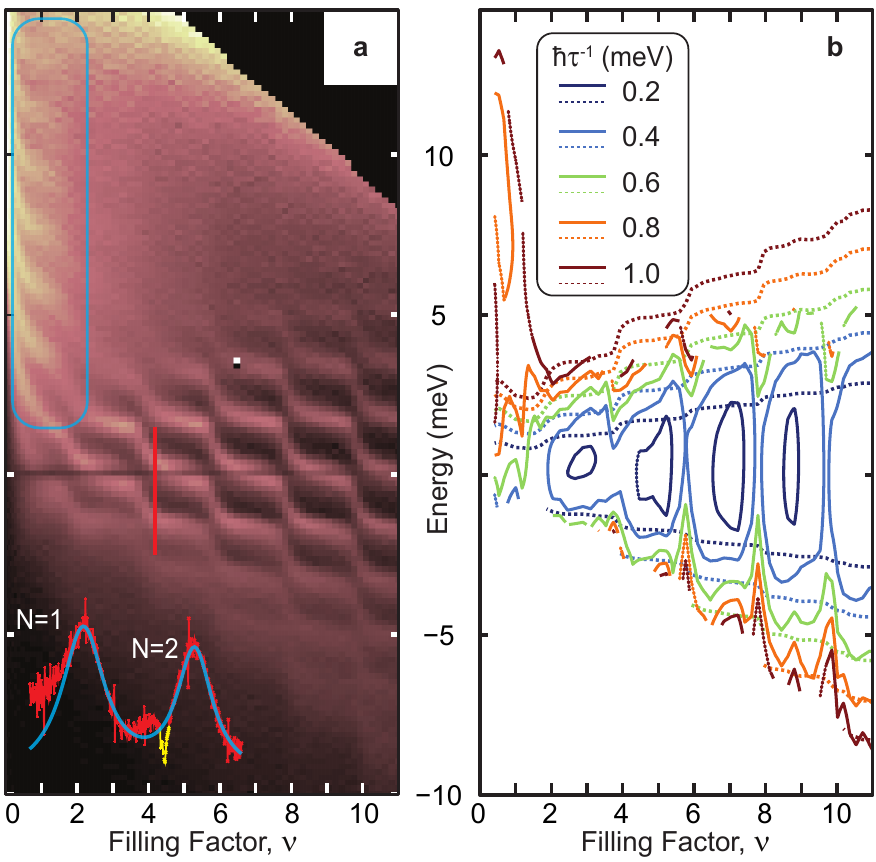}}
\caption{{\bf Quasiparticle lifetimes.} {\bf a}, TDCS spectra at 1T.
At low densities all Landau levels in the spectrum
are resolved (blue box), while at higher densities the short lifetime
due to electron-electron interactions broadens the Landau levels
farthest from the Fermi surface until they are not resolved.  A line
cut taken through the red line in the main plot with temperature
limited resolution and fit by a superposition of two Lorentzians is
shown as an inset.  The Coulomb gap is highlighted in yellow in this
line cut.  Deviations of the fit at large negative energies (left) are
from the tail of the N=0 Landau level, not included in the fit.  {\bf b}, Solid
lines; contours of constant Landau level HWHM, as obtained by fitting the
peaks in {\bf{}a} with Lorentzians and interpolating between peaks with
a fourth-order polynomial.  Dashed lines; $\hbar/\tau_{ee}$ from Chaplik's
result\cite{Chaplik71a} with no adjustable
parameters or scaling (see supplement).}
\label{fig:lifetime}
\end{figure}
\end{cbunit}

\textbf{Supplementary Information}

\begin{cbunit}
\newcommand{\mV}{\mathrm{V}}
\newcommand{\mE}{\mathrm{E}}
\newcommand{\mP}{\mathrm{P}}
\renewcommand{\figurename}{Supplementary Figure}
\hyphenpenalty=1000
\tolerance=1000
\suppressfloats[t]
\noindent\textbf{Matrix element correction}

Here we describe a physically motivated smooth transformation applied to the
SPDOS (color scale) of Figures 2 and 4 that results in the normalization
of contrast across each spectrum and the removal of an offset from the SPDOS.
This transformation only alters the contrast in our spectra and does not
change or distort the density ($x$) or energy ($y$) axes in any way.

In our experiments, we determine the SPDOS of our 2DES by
measuring the tunnel conductivity, $dI/d\mV$.  This conductivity can
be calculated using Fermi's golden rule\cite{Smoliner96}:
\begin{equation}
\label{Ieqn}
  I \propto \int_0^{e\mV} \left|M\right|^2 \rho_{3D}\left(\varepsilon\right)\rho_{2D}\left(\varepsilon\right) d\varepsilon
\end{equation}

Here $\rho_{3D}$ is the initial SPDOS in our electrode, $\rho_{2D}$ is
the final SPDOS in the 2DES, $M$ is the tunneling matrix element
connecting them, and $V$ is the voltage between the 2DES and the tunneling
electrode.  In many tunneling experiments\cite{Giaever60},
measurements are performed over a sufficiently narrow range of
energies that $\rho_{3D}$ and $M$ are approximately constant.  In this
case $dI/d\mV$ is proportional to $\rho_{2D}$, the SPDOS.  However,
because the range of tunneling energies $e\mV$ in our measurement is
significant compared to the height of our tunnel barrier and the Fermi
energy of our 3D electrode, both $M$ and $\rho_{3D}$ vary within our
spectra.  Thus, $dI/d\mV$ is no longer exactly proportional to
$\rho_{2D}$.  The changing $M$ and $\rho_{3D}$ give rise to both a
smoothly changing scaling of the SPDOS (color axis) in our spectra and
a smoothly changing offset.  Both can be seen in the $dI/d\mV$
spectrum in Suppl. Figure \ref{mat_comp}a.  Because the variation in
the scale and offset is small across the width of any given peak in
our spectra, they do not measurably affect the observed width or
position of any of the Landau levels.  While all of the information in
the TDCS spectra is present and accessible in the raw $dI/d\mV$ data,
even approximate compensation for these variations simplifies visual
inspection and fitting of the spectra.  Our strategy in performing
this compensation is first to approximate $M$ and $\rho_{3D}$ using
first principles calculations and then to ``correct'' the current to
the value it would have if $M$ and $\rho_{3D}$ were not changing.
By applying this correction before differentiating $I$, we remove both
the scaling and offset from our spectra.  For brevity, we refer to
this correction as a ``matrix element correction'' although it
includes terms from both the variation of the tunnel matrix element
and the 3D density of states.

We begin by developing an a~priori estimate of $M$ and $\rho_{3D}$ for
non-interacting electrons in zero magnetic field.  This estimate is a
generalization of the approach in [29] to non-equilibrium
tunneling.  Equating the initial and final energies during tunneling,
we have
\begin{equation}
\frac{\hbar^2}{2m^*}\left(k_{2D,x}^2 + k_{2D,y}^2\right) + \mE_0 = \frac{\hbar^2}{2m^*}\left(k_{3D,x}^2 + k_{3D,y}^2 + k_{3D,z}^2\right)
\end{equation}
Here, $k_{2D}$ and $k_{3D}$ are the wavevectors in the 2DES and 3D electrode
respectively with the $z$ direction oriented perpendicular to the plane of the
electrons in the 2DES. $\mE_0$ is the energy difference between the $k=0$
states in the 3D and the 2DES, as shown in Suppl. Figure \ref{band}.  Because our structure
has translational symmetry parallel to the quantum well, in-plane
momentum is conserved in tunneling\cite{Eisenstein91}.  In this case,
$k_{2D,xy} = k_{3D,xy} \equiv k_{xy}$. Then
$k_{3D,z}=\frac{2m^*}{\hbar^2}\sqrt{\mE_0}$ is determined entirely by
the geometry of the structure and the applied voltage, and it is the same
for all electrons able to tunnel.

We can represent this condition graphically (see Suppl. Figure
\ref{fermi_sphere}).  First, we plot the 3D Fermi sphere in $k$ space.
We then plot the 2D Fermi disc at
$k_z=\sqrt{\frac{2m^*\mE_0}{\hbar^2}}$, ensuring that states in the
2DES are plotted at the same $k_x$,$k_y$, and $k_z$ as the state in
the 3D electrode into which occupying electrons can tunnel while
conserving energy and momentum.  At equilibrium, Suppl. Figure
\ref{fermi_sphere}a, the 2D Fermi disc and the 3D Fermi sphere are
filled out to the same circle in $k$ space.  Upon disequilibrating the
two by applying a pulse, changing $\mE_0$ and thus the value of $k_z$
at which we plot the 2DES, there is a annulus of states which are now
occupied in the 2DES but not in the 3D electrode, or vice-versa.
These states, depicted in green in Suppl. Figure
\ref{fermi_sphere}b-f, contain electrons in either the 2DES or 3D
electrode that are able to tunnel.

The 1D $z$ tunneling problem separates from the plane wave states in
the $x-y$ plane\cite{Smoliner96}, so the result only depends on the
bound state wavefunction in the $z$ direction and $k_{3D,z}$ (not on
$k_{xy}$); the value of $k_{3D,z}$ is the same for all electrons able
to tunnel.  Within a given structure, we can parameterize these using
$\mE_0$.  Similarly, because the $x-y$ momentum is fixed by momentum
conservation, the 3D density of states $\rho_{3D}$ is reduced to a 1D
density of states depending only on $k_{3D,z}$, which in turn only
depends on $\mE_0$.  This allows us to move these terms out of the
integral in \ref{Ieqn}, giving
\begin{equation}
  I \propto \left|M(\mE_0)\right|^2 \rho_{3D}\left(\mE_0\right) \int_0^{e\mV} \rho_{2D}\left(\varepsilon\right) d\varepsilon
\end{equation}

Because these terms lie outside the integral, we can correct our
tunnel current simply by dividing our measured current by our
estimate for $\rho_{3D} |M|^2$.  We compute $M$ using the
transfer matrix formalism\cite{Bardeen61} and the WKB\cite{Liboff} approximation.
In calculating $M$, there is a leading factor of $k_{3D,z}$ that
arises because electrons with large momenta towards the barrier
penetrate more deeply.  There are also several smaller corrections from the change
in the shape of the barrier and the polarization of the quantum well
by the applied voltage.  It is convenient to include these by expanding
 $M$ as $M=k_{3D,z} M_0 (1-\mV_\mP/\mV_m)$, where $\mV_m$ is a constant
calculated from the first order variation of the tunneling matrix
element $M$ and $\mV_\mP$ is our applied pulse height in
volts.  The remaining 1D portion of the 3D density states is
proportional to $1/k_{3D,z}$, so $\rho_{3D} |M|^2$ is proportional to
$k_{3D,z} |M_0|^2 (1-\mV_\mP/\mV_m)^2$.  $k_{3D,z}$ is
proportional to
$\sqrt{(\mV_\mathrm{align}-\mV_\mP)/\mV_s}$, where
$\mV_{align}$ is the applied voltage pulse that sets $\mE_0=0$, exactly
aligning the $k=0$ states of the 2D and 3D bands, and $\mV_s$ is a
scale factor.  We then
``correct'' our current for variation of $M$ and $\rho_{3D}$ as follows:
\begin{equation}
\label{coreq}
   I' = I \frac{\sqrt{V_s}}{\sqrt{\mV_\mathrm{align}-\mV_\mP}} \frac{1}{\left(1 - \mV_\mP/\mV_m\right)^2}
\end{equation}
Here $I'$ is the corrected current, and $I$ is the measured current.
We estimate $\mV_\mathrm{align}$, which is a function of the 3D Fermi
energy and the 2D density, by fitting calculated tunneling I-V curves
to our zero field data at two densities and linearly extrapolating to
all densities.  The constant $\mV_s$ gives an overall scale to the
current and can be neglected, while $\mV_m$ is a structure-dependent
constant, calculated using WKB to be approximately 0.19
Volts for the structure used in this letter.

Because the in-plane momentum is conserved, the tunneling current
vanishes if we attempt to inject an electron into a state in the 2DES
with a momentum larger than the Fermi momentum in the 3D electrode.
This happens when $\mV_\mP > \mV_\mathrm{align}$, corresponding to
applied voltages slightly larger than that depicted in Suppl. Figure
\ref{fermi_sphere}f.  In this region $\rho_{3D}$ is zero, and our
correction diverges ($\sqrt{\mV_\mathrm{align}-\mV_\mP}$ in the
denominator of Equation \ref{coreq} goes to zero) because no tunneling can
occur that conserves momentum and energy.  As we approach this
divergence, we are multiplying our experimental data by increasingly
large numbers.  Well before the correction actually diverges, our
SPDOS spectra become dominated by experimental noise.  We truncate the
data near the point where noise begins to dominate to remove this
distraction and substitute a black band.  This band is visible in the
top-right (high density, high energy) corner of the SPDOS spectra in
Suppl. Figure \ref{mat_comp}b and those in Figures 2 and 4.  Near
these black bands the sharp increase in the corrected SPDOS visible in
these figures is an artifact from our rough
measurement of $\mV_\mathrm{align}$ and low order approximation of
$M$.

This correction is not exact.  Not only is $M$ calculated to low
order, but $\mV_\mathrm{align}$ is only roughly estimated, and we have
neglected the effects of interactions.  However, the correction does
serve to normalize contrast and remove offsets throughout our spectra,
simplifying both fitting and visual inspection. Suppl. Figure
\ref{mat_comp} shows a sample spectrum before and after this matrix
element correction; all of the same peaks and features are visible and
at the same places; they are simply easier to see after the
correction.  The observation that the Landau levels all have
approximately equal spectral weight after the correction suggests that
we have successfully removed the main effects of the tunneling matrix element and 3D
density of states variation from the spectra.

\noindent\textbf{Theoretical electron-electron scattering lifetime}

In Figure 4 of the letter, we plot our experimentally extracted line
widths against the expected theoretical line width due to electron-electron
scattering, as first calculated by Chaplik\cite{Chaplik71a}.  Chaplik's
result, also derived by Giuliani and Quinn\cite{Giuliani82}, is given
in terms of quantities used in this letter by:
\begin{equation}
  \frac{\hbar}{\tau_{ee}}\left(E\right) \simeq \frac{E_f}{4 \pi} \left( \frac{E}{E_f}\right)^2 \left[\ln\left(\frac{E_f}{E}\right)+\frac{1}{2}+\ln\left(\frac{2 q_{TF}}{k_f}\right)\right]
\label{chaplik}
\end{equation}
Here $q_{TF}=2me^2/\hbar^2$ is the Thomas-Fermi screening wave vector
in 2D, $k_f$ is the Fermi wavevector, and $E_f$ is the Fermi energy.
We extract $E_f$ from the experimental data by measuring the energy of
the lowest ($N=0$) Landau level relative to the Fermi surface ($E=0$
on the spectrum in Figure 4a), and approximate $k_f$ as $\sqrt{2 m
E_f}/\hbar$.  This leaves no free parameters in Chaplik's result,
allowing us to make a direct comparison with our data.
\bibliographystyle{naturemag_2} \bibliography{article}

\begin{figure}
\center\resizebox{6in}{!}{\includegraphics{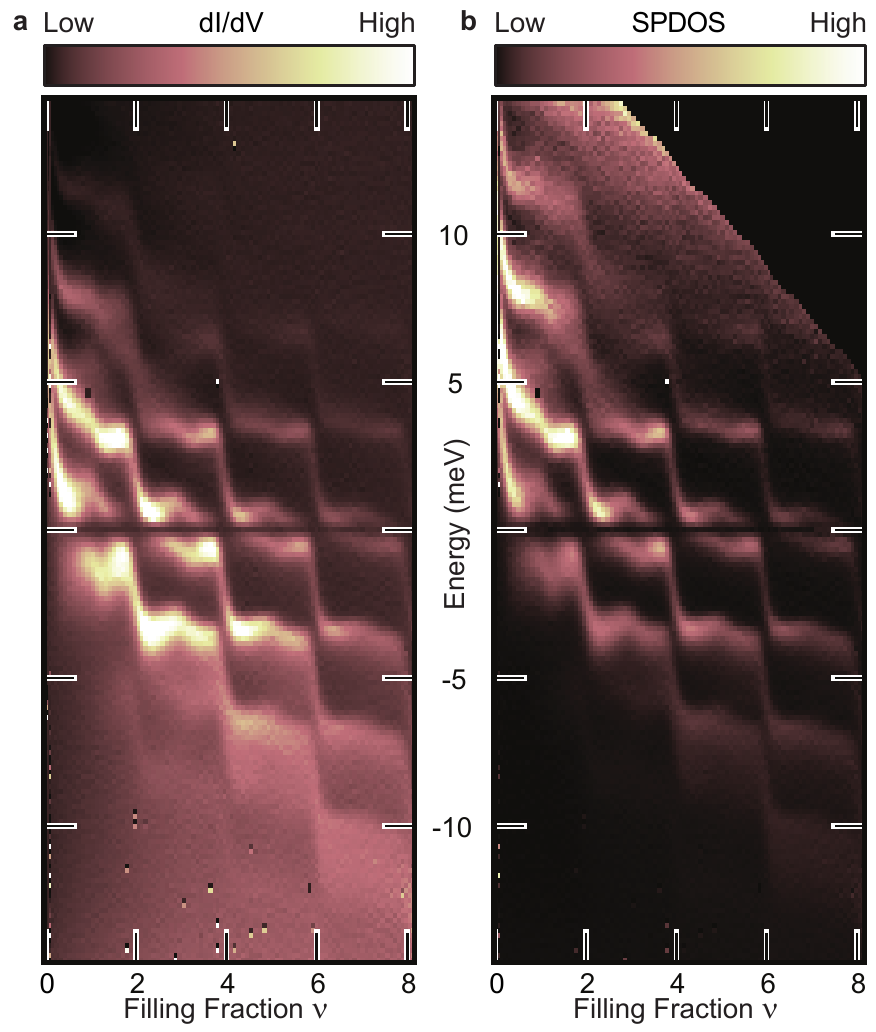}}
\caption{\textsf{\textbf{dI/dV and SPDOS spectra at 2 Tesla.}}  Application of
the matrix element correction to the dI/dV spectra in \textbf{a} yields the
SPDOS spectrum shown in \textbf{b}.  Note
the positions of peaks and peak widths are unchanged, but the
corrected version (SPDOS) has more uniform contrast across the peaks and a smaller
background.
 \label{mat_comp}}
\end{figure}

\begin{figure}
\center\resizebox{6in}{!}{\includegraphics{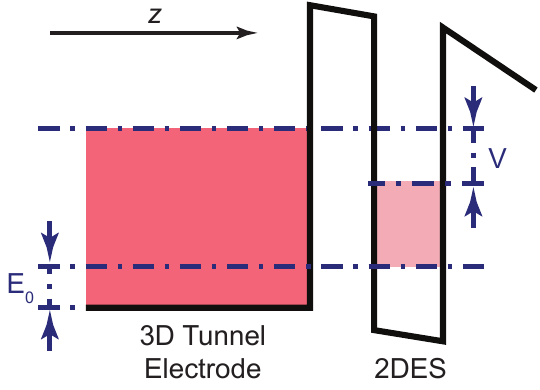}}
\caption{\textsf{\textbf{Definition of $\mE_0$.}} Band diagram of the
quantum well and tunneling electrode, showing $\mE_0$, the energy
difference between the quantum well bound state energy and the bottom
of the conduction band in the 3D electrode.  $\mE_0$ depends on any
applied AC and DC voltages.
\label{band}}
\end{figure}

\begin{figure}
\center\resizebox{6in}{!}{\includegraphics{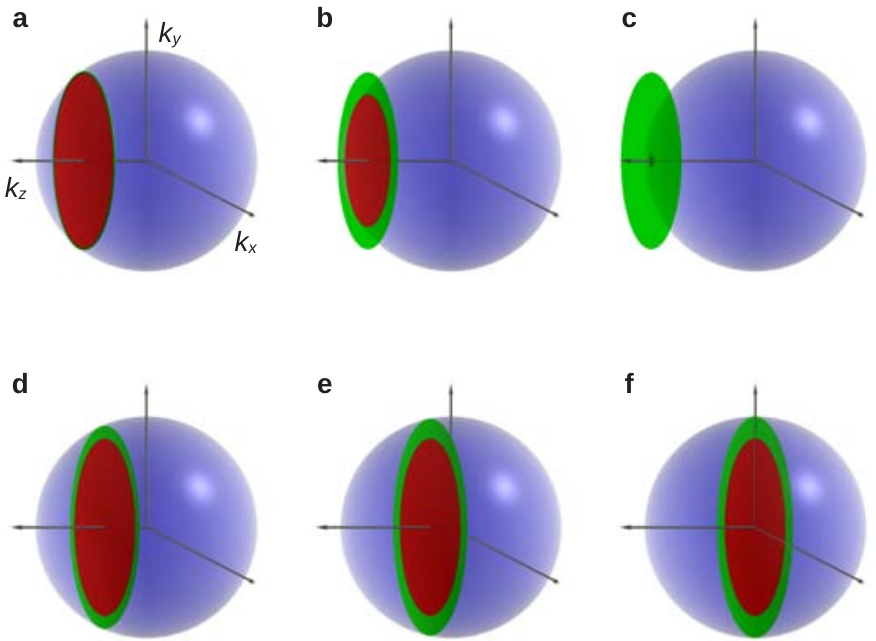}}
\caption{\textsf{\textbf{Momentum conservation in tunneling.}}  The 3D
Fermi sphere and 2D Fermi disc are shown as a function of $k$ vector
with a fixed density in the 2DES but with a variety of voltages
applied between the 2DES and the Fermi sphere.  The Fermi disc has
been positioned at the value of $k_{3D,z}$ into which electrons tunnel
from the 2DES while conserving transverse momentum and energy.
\textbf{a,} shows the momentum space representation when the 3D Fermi
sphere (blue) and 2D Fermi disc (red) are at equilibrium.  Only the
electrons in the ring of states (green) at the intersection of the
Fermi sphere and Fermi disc are able to tunnel.  \textbf{b,} shows the
Fermi disc and sphere after a pulse has been applied to eject
electrons from the 2DES.  Electrons in the annulus of filled states in
the Fermi disc outside of the Fermi sphere are now able to tunnel.
These all tunnel into states with the same $k_{3D,z}$.  \textbf{c,}
shows the situation where a sufficiently large pulse has been applied
to allow electrons in all the states in the 2DES to tunnel. This
situation corresponds to tunneling into the lowest lying (most
negative energy) Landau level in our spectra.  \textbf{d-e,} show two
small injecting pulses, where electrons will be injected into the
annulus of states (green) that were once empty in the 2DES but now lie
inside the Fermi sphere.  \textbf{f,} shows a large injecting voltage.
If the injecting voltage is increased any further the tunnel current
will be cut off because $\mV_\mP > \mV_\mathrm{align}$ (see text).
This condition corresponds to the beginning of the black bands in the
top-right corner of our SPDOS spectra.
 \label{fermi_sphere}}
\end{figure}

\end{cbunit}
\end{document}